\documentclass[11pt]{article}
\usepackage{paspconf}

\usepackage{astrobib}
\usepackage{epsf}
\def \beginfig          {\begin{figure}}
\def \endfig            {\end{figure}}

\def \begineq           {\begin{equation}}
\def \endeq             {\end{equation}}

\def \d {\partial}

\def\gtorder{\mathrel{\raise.3ex\hbox{$>$}\mkern-14mu
             \lower0.6ex\hbox{$\sim$}}}
\def\ltorder{\mathrel{\raise.3ex\hbox{$<$}\mkern-14mu
             \lower0.6ex\hbox{$\sim$}}}

\def \hide#1{}

\def\br {{\bf r}}

\def\aj{{AJ}}                   
\def\araa{{ARA\&A}}             
\def\apj{{ApJ}}                 
\def\aap{{A\&A}}                
\def\pasp{{PASP}}               

\newcommand{\astrophref}[1]{{\tt http://xxx.lanl.gov/abs/astro-ph/#1}}

\newcommand{\href}[2]{{\tt #2}}

\begin{document}

\title{Weak Lensing by Galaxy Clusters}
\author{Nick Kaiser}
\affil{Institute for Astronomy, University of Hawaii, 2680 Woodlawn Drive, Honolulu, HI 96816.
\href{http://www.ifa.hawaii.edu/~kaiser}{{\tt http://www.ifa.hawaii.edu/$\sim$kaiser}}}

\begin{abstract}
In this talk I review some of the key questions that weak lensing
observations of clusters can potentially answer, and sketch the
progress that has been made to date in extracting quantitative
estimates of masses and density profiles.  A major difficulty
in interpreting these measurements is the diverse
methods that have been used to calibrate the shear-polarization
relation, none of which are completely satisfactory.  
I describe some recent developments in the 
theory of weak lensing calibration.  
I then present some results from the UH8K
camera and discuss their implication for cosmology.  Finally, I outline
a new strategy for obtaining high resolution optical imaging
from the ground using an array of small telescopes, and describe
how such an instrument can greatly improve the power of weak lensing
observations.
\end{abstract}

\section{Introduction}

Galaxy clusters act as gravitational lenses and distort the shapes and
sizes of faint background galaxies.  One can distinguish two regimes:
the central strong lensing regime characterized by more or less obviously distorted
galaxies, and the weak lensing regime at larger impact parameters where
the distortion is weak and must be measured by averaging shapes of many galaxies.
The division between these regimes lies at around $100 h^{-1}$kpc
for a typical massive cluster. In this
talk I shall focus on the weak lensing regime.  From the shape distortion
or `image shear' one can  in principle recover the full
2-dimensional projected total mass density, and over a wide range of
scales, making this a  very  direct and powerful probe of the dark matter.
Mass determinations to date have been somewhat noisy --- the
precision in the mass being limited by the finite number density of
faint background galaxies --- but coarse properties such as the 
mass profile can be detected at typically the 4-10 sigma level.

Weak lensing measurement of clusters can address a number of key questions
in cosmology.  First, it is of some interest to compare the lensing
mass measurements with independent probes such as virial analysis and
X-ray temperatures, since this provides a test of the various simplifying
assumptions typically used in these techniques such as dynamical
equilibrium, orbital anisotropies, spherical symmetry etc.
Second, since lensing measures the projected mass, and the same observations
yield the projected light of the lensing system, it makes sense to form the
mass-to-light ratio.  Questions which one might then hope to 
answer include: Is $M/L$ universal? Does $M/L$ vary with radius within
individual clusters? Does $M/L$ vary with the lens redshift?
Are there `dark clumps' out there which have so far eluded detection?
This analysis should directly reveal any `biasing', or differences
between the mass profile and the profile traced by the various types of
galaxies, and the lensing approach is unique in being applicable
seamlessly in both the equilibrium and infall regimes.  It
should also yield important constraints on the evolution of
massive clusters, again with obvious implications for cosmological structure
formation theory.  Next, though this tack is somewhat more model
dependent, one can compare the total mass with the X-ray derived
gas mass, and with the total baryonic mass (gas plus baryons in galaxies).
Again, one would very much like to know whether the 
$M_{\rm baryon} / M_{\rm total}$ ratio is universal, whether this ratio varies
with radius within clusters and whether there are baryon poor
or high gas entropy clusters.  Finally, there is the question of the
redshift distribution of the faint background galaxies, which enters as
a calibration factor in weak lensing mass determinations. 
The variation of the mean redshift, or more generally the redshift
distribution, as a function of magnitude say can provide important
clues for galaxy formation. Finally,
distortion strength depends on the ratio of 
lens-source to observer-source distances $D_{\rm LS} / D_{\rm OS}$. 
It is therefore of interest to compare lensing derived distances
with spectroscopic redshifts, since the redshift-distance relation
can be used to constrain cosmological parameters $\Omega_{\rm matter}$, 
$\Omega_\lambda$ etc. 

Weak lensing observations have now been made for
around 30 clusters, and the status as of early 1999
has been extensively reviewed by \cite{mellier99}.
He gives a very useful distillation of the data as
a table of mass-to-light ratios, along with lens redshift, impact parameter
for the observations etc. The $M/L$ values are typically a few
hundred (times $h$), not so different from those obtained 
from virial and X-ray studies.  There is however considerable
variation from cluster to cluster with extreme values of $M/L$ ranging
from $\sim 100h$ to $\sim 1000h$.
There is also an indication of an increase in $M/L$ with radius in at least
one case. If correct, these are results of profound significance
but it is difficult to know exactly what to make of this.  Not all of the entries
in the table have rigorous error estimates; different observers 
have quoted the $M/L$ for different passbands; different assumptions
have been made about the background galaxy redshifts;
different models have been used to correct for evolution of
the cluster galaxy population, and different techniques have been
used to calibrate the shear-polarization relation.  
Put together,   
these problems make it very hard to draw any definitive conclusions
about systematic $M/L$ variations.
The collection of results to date have shown
that the technique is certainly viable, and have convincingly dispelled 
fears that the measurements are
contaminated by instrumental effects etc. I would argue that
the major `nuisance factors' listed above are now quite well understood
and the time is ripe for a significant advance in
well calibrated and quantitative mass measurements which
can convincingly confirm or refute the suggestions of mass-to-light variations
from the current data.

In the rest of this talk I shall first describe some advances in
our understanding of how to convert shape polarization measurements to
quantitative shear estimates.  I will then review an application to
a redshift 0.4 supercluster using the UH8K camera, and the intriguing
relation between the mass and the light in this system.
Finally I describe a strategy for deep wide field optical imaging
using an array of small telescopes with on-chip fast guiding, and 
the gains in improvement in lensing observations afforded by
this technique.

\section{Calibration of the Shear Polarization Relation}
\label{sec:calibration}

\def \obs       {{\rm obs}}
\def \fobs      {f_{\rm o}}
\def \fint      {f}
\def \fs        {f_{\rm s}}
\def \fcrit     {f_{\rm crit}}

\def \gdag      {g^\dagger}

\def \d {\partial}

\def \Maij {M_{\alpha i j}}
\def \Mbij {M_{\beta i j}}
\def \Malm {M_{\alpha l m}}
\def \Mblm {M_{\beta l m}}
\def \Q {{\cal Q}}
\def \K {{\cal K}}
\def \P {{\cal P}}
\def \R {{\cal R}}

\def \Peff {{\overline P}}

The effect of a weak gravitational lens is a mapping of the surface brightness $f$
of distant objects:
\begineq
\label{eq:lensmapping}
f'(r_i) = f((\delta_{ij} - \psi_{ij}) r_j)
\endeq
where $r_i$ is the angular position on the sky
and $\psi_{ij}$ is the symmetric `distortion tensor' 
which is an integral along the line of sight of the
transverse components of the tidal field:
\begineq
\label{eq:psifromphi}
\psi_{lm} = 2 \int d \omega {\sinh \omega \sinh (\omega_s - \omega) \over 
\sinh \omega_s} \partial_l \partial_m \Phi
\endeq  
\cite{gunn67}, (and see more recent discussions 
\citeN{barkana96}, \citeN{k98})
where $\omega$ is conformal distance and the
potential $\Phi$ is related to the density contrast by 
$\nabla^2 \Phi = 4 \pi G \delta \rho$.  
These results can be generalized to deal
with sources at a range of distances, and with either accurately
known redshifts or partial
redshift information from broadband colors, and provide
a direct quantitative connection between the observable distortion and the total mass
density. What has proved more problematic has been establishing a precise
relation between the distortion and the shape polarization actually measured,
as I shall now review.

Specializing to a trace-free
distortion tensor $\psi_{ij} = \{\{\gamma_1, \gamma_2\}, \{\gamma_2, -\gamma_1\}\}$
the mapping (\ref{eq:lensmapping}) can be written as 
\begineq
f' = S_\gamma f
\endeq
where the `linearized shear operator' is 
\begineq
\label{eq:linearshearoperator}
S_\gamma = 1 - \gamma_\alpha \Maij r_i \partial_j
\endeq
and where we have introduced the constant $2 \times 2$
matrices $M_0 = \{\{1,0\},\{0,1\}\}$,
$M_1 = \{\{1,0\},\{0,-1\}\}$,
$M_2 = \{\{0,1\},\{1,0\}\}$.  
This operator is similar to a rotation operator
and provides a convenient way to compute the 
effect of a shear on measurable shape statistics. 

It was realized early on \cite{vjt83} that
a natural way to measure the shear is to look for a
non-zero systematic average of the 
second central moments of the background galaxy images.
\begineq
\label{eq:secondmomentdefinition}
q_{lm} = \int d^2 r \; r_l r_m f(r) .
\endeq
Indeed, if we define $q_A = {1 \over 2} M_{Aij} q_{ij}$
(so $q_0$ is a measure of the size
of the object and $q_\alpha$, $\alpha = 1,2$ 
is the `shape polarization') and apply the shear operator
to compute the response of the $q_A$, one finds that  
a fair estimate of the
shear is
\begineq
\label{eq:simpleshearestimate}
\hat \gamma_\alpha = \langle q_\alpha \rangle / 2 \langle q_0 \rangle
\endeq
This simple estimator is less than ideal in that the
averages of $q_\alpha$ and $q_0$ are heavily
weighted towards larger galaxies, but illustrates the general idea
and the approach can be
generalized to give a shear estimator in terms of
the `ellipticity vector' $e_\alpha = q_\alpha / q_0$
(\citeNP{bm95}; \citeNP{ksb95} hereafter KSB).
A more serious shortcoming of (\ref{eq:simpleshearestimate})
is that it ignores the effect of atmospheric `seeing'
and instrumental point spread function (PSF) $g(r)$.  
The combined effect of gravitational shearing and the PSF convolution is
to give an observed surface brightness
\begineq
\label{eq:fobsprimedefinition}
\fobs' = g \otimes S_\gamma f .
\endeq
Circular seeing tends to reduces the ellipticity
while departures from circularity will introduce an artificial
polarization. We need some way
to correct for the latter and calibrate the
effect of the former.  Interestingly, 
for moments defined as in (\ref{eq:secondmomentdefinition})
this is trivial since under convolution these
moments add \cite{vjt83}
so one can correct the averaged moments appearing in (\ref{eq:simpleshearestimate})
by simply subtracting the moments for stellar objects.

Unfortunately, the moments as defined in (\ref{eq:secondmomentdefinition}) are
useless in practice for a number of reasons; noise from photon counting and
from other objects seen close by on the sky diverges strongly with
the integration limit, and there is also the serious problem
that for realistic PSFs the second moment itself does not converge.
What has been done in practice is to truncate the second moment
integral either with an isophotal threshold 
as in the FOCAS software
package \cite{jt81} and also the Sextractor package
\cite{ba96}
or by introducing some kind of user defined weight function $w(r)$
in (\ref{eq:secondmomentdefinition})
and computing shapes in terms of the weighted moments
(\citeN{bm95}, KSB).
In either case the addition law for the moments
no longer holds, and compensating for the effects
of the PSF becomes considerably more complicated.

A partial solution to this problem was given in KSB
(see also \citeN{hfks98})
who computed the response of ellipticities
to an anisotropy of the PSF under that assumption that this
is a convolution of a circular PSF $g_{\rm circ}(r)$ with some compact 
but highly anisotropic function $k(\br)$.  They found a response
proportional to polarization computed from the unweighted second moments of $k(\br)$.
They also computed the response of ellipticities
formed from weighted moments to a shear applied to $\fobs$
{\sl after\/} smearing with the PSF. This of course is
not what one wants, as one really needs the response to a shear applied
before smearing with the PSF.
\citeN{lk97} extended the KSB analysis to account for
finite resolution, but under the restrictive
assumption that the PSF has a Gaussian profile. 
\citeN{rfg99} have also attempted to generalize the KSB approach
to properly allow for finite resolution.  
There is a serious problem with these approaches:  The `convolution model' of
KSB and subsequent extensions is too restrictive.  
It is a good approximation for the simple case of atmospheric seeing
plus small amplitude guiding errors or aberrations, but the
perturbation expansion, of which the KSB result is the lowest order
term, breaks down for large errors. For diffraction limited observations matters are
worse as the unweighted central moment which appears here 
is dominated by the extreme wings of the 
PSF, so applying the KSB formalism then makes no sense at all.
The KSB analysis is something of a blind alley and 
a fundamentally different approach is needed.

To make progress we need to return to (\ref{eq:fobsprimedefinition}).
Using the convolution theorem one can
rewrite this as
\begineq
\label{eq:fobsoperator}
\fobs' = \fobs - \gamma_\alpha \Maij 
(r_i \d_j \fobs - (r_i \d_j h) \otimes \fobs).
\endeq
\cite{kaiser99}
where $h$ is the inverse  transform of
$\tilde h \equiv \ln \tilde g$, i.e.~the logarithm of the optical transfer function 
(OTF).
This `finite resolution linearized shear operator' is extremely powerful
as it gives the response of an image to a shear applied before seeing purely
in terms of observable quantities. It
is the regular
shear operator $S_\gamma$ applied to the post-seeing image $\fobs$ plus
a commutator term
which is a correction for finite PSF size and which is a
convolution of $\fobs$ with a kernel  function
$\gamma_\alpha \Maij r_i \d_j h$.
For the Gaussian PSF model the commutator term consists of
second derivatives of $\fobs$, and so in this case the shear
operator is a local differential operator.  This is a very special
case.  
In the case of atmospheric turbulence limited seeing 
the log-OTF transform has a power law form $h(r) \propto r^{-11/3}$, 
so then commutator term is highly non-local and qualitatively
different from the Gaussian model. 

For
diffraction limited observations things are a little more problematic,
as the log-OTF $\tilde h$ then diverges as one approaches the diffraction
limit. However, the amount of information in these marginal frequencies
is asymptotically zero, and one can remove the divergence by computing
shapes from an image $f_s$ which has had the marginally detectable modes
attenuated.
One simple option is to re-convolve with the
PSF itself. This approach might seem surprising since weak lensing observations
tend to be `resolution starved', and the obtainable signal to noise is a strong function
of the seeing width.  However, it is important to realize that unlike bad seeing,
the convolution
here is applied {\sl after} the photon counting 
noise is realized in the CCD detector, so there is consequently no 
loss of information.
In this case response to a shear is
\begineq
\label{eq:fsoperator}
\fs' = \fs + \gamma_\alpha \Maij 
(2 (r_i \d_j g) \otimes \fobs -  r_i (\d_j g \otimes \fobs)) .
\endeq
A nice feature of this approach is that with a minor modification one can
at the same time `null out' any instrumental shape anisotropy.
Equations (\ref{eq:fobsoperator}), (\ref{eq:fsoperator}) are quite general and 
allow one to predict the
response of the shape statistic for an individual object to a
shear, much as was down by KSB.  They are slightly more complicated to implement
since they require the full 2-dimensional form of the PSF rather than
just the second moments, but this turns out not to be too difficult in
practice.  

The individual object response computed in this way could be used to
construct a fair shear estimator simply by averaging polarizations and
dividing by the average polarizability much as above.
However, as before, this unweighted
averaging is not ideal as the information content is strongly dependent on
the size, brightness and eccentricity of the objects.  To make an
optimally weighted shear estimator we need to know the conditional
mean polarization for galaxies of given
flux, size and shape.  This leads to an `effective polarizability'
which allows one to properly combine estimates of the shear from
different types of galaxies. 
The great advantage of this approach is that for the first time
it allows one to construct an optimized minimum variance weighting scheme
for combining shear estimates and allows one to define a useful
`figure of merit' --- essentially an inverse variance per steradian -
to quantify the power of any given data for measuring 
gravitational shear.
For example, a 2.75hr I-band imaging at CFHT
gave $2.85 \times 10^5 / {\rm sq\; degree}$.
These relations allow one to
to objectively  tune the parameters of one's shape measurement
scheme in an unbiased and objective
manner, and can be modified to include e.g.~photometric photometric information.

\section{MS0302}

We have applied this analysis to deep multicolor images of MS0302 taken with the
UH8K camera at the CFHT.  The field contains three physically associated
massive clusters at $z \simeq 0.42$ \cite{kwl+98}.  Shear analysis of the
30,000 or so faint galaxies in the field reveals three major mass concentrations
coincident with the optical and X-ray locations.  There is also evidence of
a further concentration which seems to be associated with a foreground
cluster at $z \simeq 0.2$.  

As discussed in the Introduction, a major goal of weak lensing cluster
studies is to compare the mass profile with that of the light
to probe galaxy `bias' directly. 
The data here are useful for this application since the
1/2 degree field allows one to probe out beyond the virial
radius of the clusters. It must be
realized here that there is no unique `light profile'; we expect
from the morphology-density relation \cite{dressler80} that the
result will be very different for early-type and late-type galaxies.
The picture for the structure of clusters that emerges from these
and other morphological studies 
is that the early-type galaxies are relatively
concentrated in the densest parts of clusters with late-type
galaxies dominating at larger radius and in the infall region.

To perform this test one cannot simply add all the bright galaxy light in the
field because of contamination by foreground galaxies.  Ideally one would
use spectroscopic redshifts to generate a prediction for the shear field
that properly incorporates galaxy distances, but unfortunately
these are not available.  Second best would be photometric
redshifts from multi-color observations.  Here we have only images
in the two passbands V and I.  This would seem to be inadequate since it
is known that at the very least galaxies form a two parameter family
(luminosity and type) so a minimum of 3 passbands is required to
determine intrinsic luminosity, type and distance.  However, what one
can do is to generate a prediction for the surface mass density
due to early type galaxies alone.  This is because the early type galaxies are
much redder than the other types.  Thus if we assign distance to a 
galaxy based on its V-I color assuming it is an early type galaxy we
will either get the right answer or, if the galaxy is really of later type,
we will grossly underestimate the distance, and therefore grossly underestimate the
luminosity, and the combination of these effects means that the
non-early type galaxies receive essentially zero weight.

\begin{figure}[htbp!]
\plotone{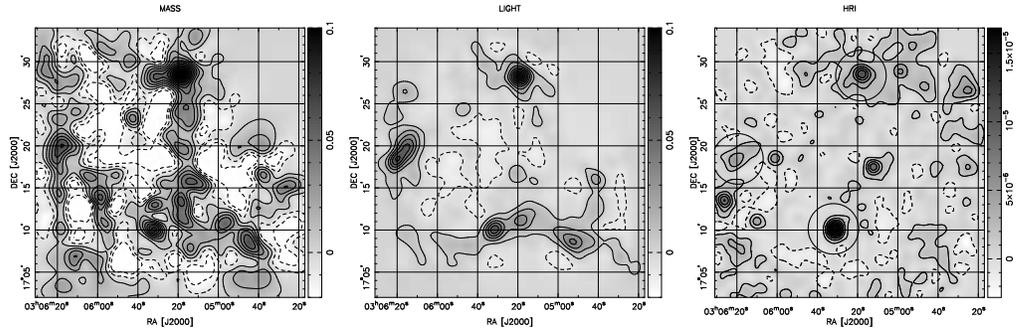}
\caption{A comparison of the mass, early-type light and X-ray distributions.
The mass reconstruction was made using the KS93 method.
The `light' image is really a prediction for the mass reconstruction
assuming early-type galaxies trace the
mass with a fixed mass-to-light ratio $M/L_B = 270 h$ in solar units.
To construct this we assign bright galaxy distances based on their color
and thus generate a prediction for the shear field and then feed this
(noiseless) shear image through the reconstruction algorithm. 
The X-ray image is from the ROSAT HRI.}
\label{fig:mlx}
\end{figure}

\begin{figure}[htbp!]
\plotone{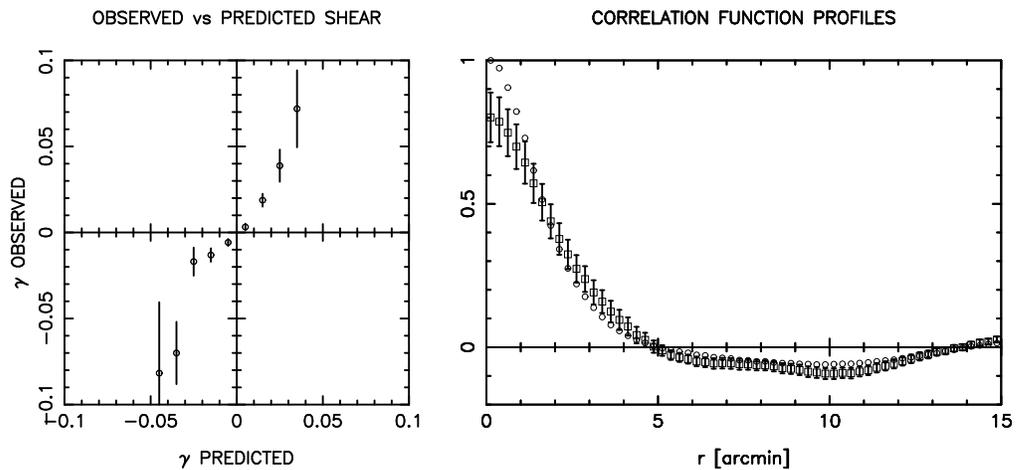}
\caption{Left hand panel shows the measured shear versus that
predicted.  Right hand panel shows the 
light-mass cross-correlation (boxes) 
and the light-light auto-correlation functions (circles).}
\label{fig:shearshearandccfprofile}
\end{figure}

A comparison of the mass computed from the faint galaxies
using the \citeN{ks93} algorithm, early-type light and X-ray distributions is shown
in figure \ref{fig:mlx}.  There seems to be good general agreement between the
mass and the early type galaxies.  This conclusion is bolstered by the
figure \ref{fig:shearshearandccfprofile} which shows on the left the
predicted versus observed shear and on the right a comparison
between the light-mass cross-correlation and the light-light auto-correlation.

These results are very intriguing.  First, the inferred mass-to-light ratio 
seems to be about a factor two lower than the value for more massive
clusters like Abell 1689 say. Second,
the mass seems to be clustered just like the early-type light.  There is
no indication that the mass profile around the clusters is more extended then the
early-type light, as might be expected on the basis of standard morphology/density
relation (though we should admit that we do not really know whether
this particular system conforms to the norm in this regard).
The physical picture that emerges is of a collection three clusters
within a roughly $6 \times 6 \times 10 (h^{-1}{\rm Mpc})^3$ cuboid.
This volume is over-luminous by about a factor 20 relative to the
spatial average, but the system as a whole is actually unbound.
Finally, if we take seriously the idea that the early-type galaxies
accurately trace the mass, and that there is
really very little mass associated with later-type galaxies,
then applying the usual accounting arguments (and
recognizing all the usual associated {\it caveats\/})
one obtains a very low value for the 
matter density of $\Omega \simeq 0.1$.  

\section{High Resolution Imaging from Small Telescopes}

Weak lensing observations from the ground are hampered by
the image degradation from the atmosphere.  Empirically
we have found from our CFHT observations that the information content
of our images (defined as the inverse shear variance per
unit solid angle) is a strongly decreasing function of the
image width.  Even at $m_I \simeq 25$ the great majority of
galaxies are poorly resolved even in excellent seeing, and
we know from the Hubble Deep Field that galaxies get
still smaller at fainter magnitudes.  In the coming era
of weak lensing observations with 8m class telescopes
this will become more and more a serious limitation.  
To get around this we have two options: either
observe from space or apply some kind of adaptive optics
correction.  Here I will focus on the latter option and
advertise a possible way to implement
wide field image correction using an array of small telescopes
\cite{ktl99}.

\begin{figure}[htbp!]
\plotone{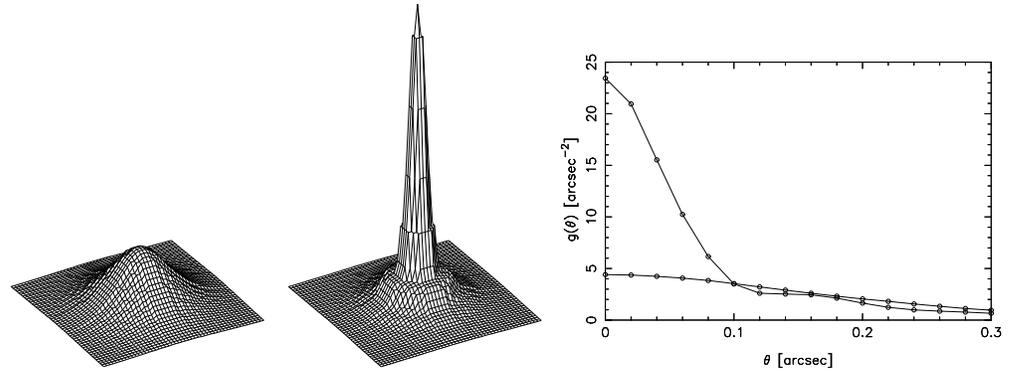}
\caption{Comparison of uncorrected natural seeing PSF for a large
telescope and fast guiding PSF for a 1.5m telescope. The FWHMs are
$0''.4$ and $0''.12$ respectively.}
\label{fig:psf}
\end{figure}

Adaptive optics systems on large telescopes can
give huge gains in angular resolution. Unfortunately they cannot
easily be applied to wide field imaging because of the
isoplanatic angle problem; objects more widely separated than
a few tens of arc-seconds sample independent paths through
the atmosphere and suffer largely independent wavefront deformations,
so to correct over a wide field requires some kind of multiplexing
of the wavefront correction method.  
However, very low order wavefront correction in the form of simple fast guiding on
small telescopes (those with aperture diameter about four times the
`Fried length' $r_0$ which characterizes the wavefront deformation;
this is $D\simeq 1.5$m at a good site) can give a
substantial gain in image quality of about a factor three in PSF
width, which means a huge gain in improvement in efficiency
for weak lensing observations.  For a telescope of this size the
instantaneous PSF is found to consist of a few diffraction limited speckles
which dance around on the focal plane,
and by tracking the brightest speckle one can obtain a 
PSF with about 30\% of the light in a tight diffraction
limited core and the rest in a halo of width similar to the uncorrected PSF.
The uncorrected {\sl vs.\/} corrected PSFs are shown in
figure \ref{fig:psf}.  

Fast guiding still suffers from isoplanatism --- images
move coherently over only about 1 arc-minute --- so for a wide field
we need to move different part of the focal plane independently.  However,
this can now be cheaply and efficiently implemented using OTCCD technology
\cite{tbs97} in which the guiding is done by moving the 
accumulating charge on the chip.
With a wafer scale device consisting of a large number of
separately addressable cells one can effectively synthesize a
`rubber focal plane'.  

\begin{figure}[htbp!]
\plotone{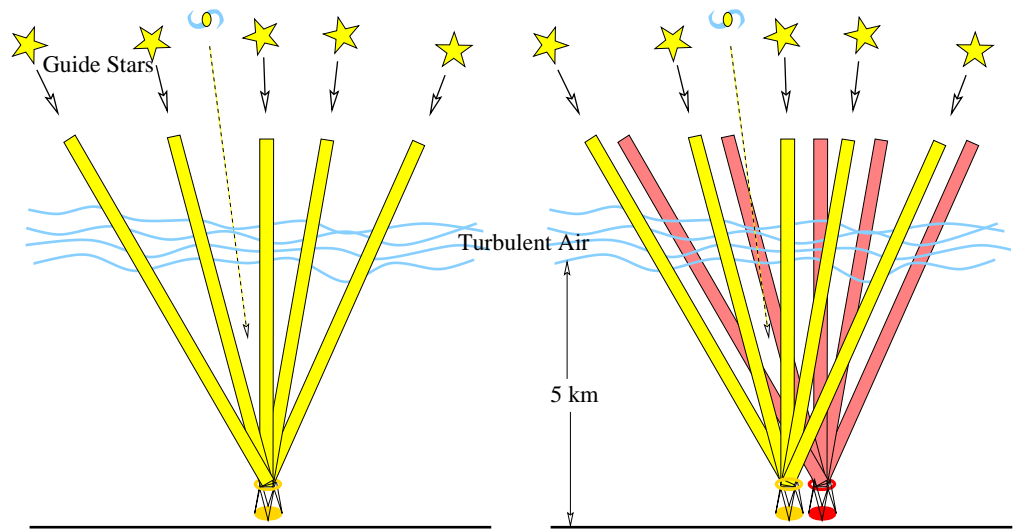}
\caption{Schematic illustration of guiding algorithm. The telescope on
the left is unable to correct for the motion of the galaxy as it lies too far from
a bright guide star.  By combining guide star information from multiple
telescopes one can fill in the missing information.}
\label{fig:beams}
\end{figure}

Some subset of the detector cells will contain
bright stars, and these can be read out rapidly to provide
the guiding information.
Even if equipped with such a device
a single telescope would not be able to provide sharp images
over the whole field of view.  This is because the surface density
of sufficiently bright guide stars falls somewhat short of that needed 
to fully constrain the deflection field.  However, with multiple
telescopes one obtains multiple samples of the atmospheric deflection
as illustrated in figure \ref{fig:beams}.
We have presented an algorithm which combines the measured deflections
from multiple telescopes; it exploits the Gaussian statistics and
highly stratified nature of atmospheric seeing and generates a
Bayesian conditional mean estimator for the deflection.  
An array of $\sim 36$ such telescopes would give the collecting
area equal to that of a conventional $9$m telescope at a comparable
cost --- the cost here scaling linearly with collecting area --- but
with greatly improved image quality.

\end{document}